\documentclass[twocolumn,longauthor]{aastex63}

\usepackage{hyperref}
\usepackage{amsmath}
\usepackage{graphicx}
\usepackage{xcolor}
\let\tablenum\relax 
\usepackage[separate-uncertainty=true,multi-part-units=single]{siunitx}
\newcommand{\dd}{\mathop{\mathrm{d}\!}{}}

\usepackage{comment}
\shorttitle{Modeling Tidal Disruptions with Dynamical Tides}
\shortauthors{Zhou, Z., ET AL.}
\graphicspath{{./}{}}

\begin{document}

\title{Modeling Tidal Disruptions with Dynamical Tides}

\author{Zihan Zhou\,$^{1}$, 
        Giovanni Maria Tomaselli\,$^{2}$, Irvin Mart\'inez-Rodr\'iguez\,$^{3}$, and Jingping Li\,$^{3}$}
\affiliation{$^{1}$Department of Physics, Princeton University,
             Princeton, NJ 08540, USA\\
             $^{2}$School of Natural Sciences, Institute for Advanced Study, Princeton, NJ 08540, USA\\
             $^{3}$Department of Physics, Carnegie Mellon University, Pittsburgh, PA 15213, USA}

\email{zihanz@princeton.edu \\
tomaselli@ias.edu \\
ifmartin@andrew.cmu.edu \\
jingping.li@aya.yale.edu}




\begin{abstract}
Tidal disruption events (TDEs) occur when stars pass close enough to supermassive black holes to be torn apart by tidal forces. Traditionally, these events are studied with computationally intensive hydrodynamical simulations. In this paper, we present a fast, physically motivated two-stage model for TDEs. In the first stage, we model the star's tidal deformation using linear stellar perturbation theory, treating the star as a collection of driven harmonic oscillators. When the tidal energy exceeds a fraction $\gamma$ of the star's gravitational binding energy (with $\gamma \sim \mathcal O(1)$), we transition to the second stage, where we model the disrupted material as free particles. The parameter $\gamma$ is determined with a one-time calibration to the critical impact parameter
obtained in hydrodynamical simulations. This method enables fast computation of the energy distribution $\dd M/\dd E$ and fallback rate $\dd M/\dd T$, while offering physical insight into the disruption process. We apply our model to MESA-generated profiles of middle-age main-sequence stars. Our code is available on \href{https://github.com/ZihanZhou26/PerTDE}{GitHub}.
\end{abstract}


\section{Introduction}\label{sec:intro}

In galactic dynamics, unlucky stars can be scattered onto orbits traveling towards the galactic center, experiencing strong tidal encounters with the central supermassive black hole (SMBH). During the encounter, the tidal force from the BH deforms the star and drives the stellar oscillations. Once the tidal force is approximately equal to the star's self-gravity, the star is torn apart and stretched into debris fluid streams. This process is famously known as a tidal disruption event (TDE) (\cite{hills1975possible,Rees:1988bf}). The stellar material that remains bound to the BH forms an accretion disk and produces an electromagnetic transient, which is the main observational signature of this event.

As famously argued by \cite{Frank:1976uy} and \cite{Magorrian:1999vm}, the TDE rate per galaxy per year is about $10^{-5} $ to $10^{-4}$, which is consistent with current observations (\cite{Donley:2002mp,vanVelzen:2017qum,Sunyaev:2021bln,Yao:2023rbr}). To date, more than $100$ TDE flares have been observed in the central regions of various galaxies, and this number is expected to increase dramatically in the next few years (\cite{Bricman:2019mcg}). Modern observations can now routinely capture both the rising and the decaying parts of the transient light curve (\cite{Holoien:2019zry,PTSSTNTS:2018dcg,vanVelzen:2018dlv,Holoien:2018oby,Hung:2020jqz}). Once the light curve is observed, we are in principle able to infer the properties of the central BH and of the disrupted stars, e.g. BH mass and spin, stellar mass and age, impact parameter, etc. This information may be used to determine the SMBH mass function and the stellar dynamics near the galactic center (\cite{Stone:2014wxa,Yao:2023rbr}). In order to extract these properties from the observed data, a great amount of hydrodynamical simulations have been conducted to study the dynamics of TDE (\cite{nolthenius1982passage,1993ApJ...418..163K,1993ApJ...418..181K,laguna1994sph,diener1997relativistic,ramirez2009star,Guillochon:2012uc,law2017low,Golightly:2019jib,goicovic2019hydrodynamical}). Pioneering simulations used polytropic stars models for simplicity, see e.g.\ \cite{Guillochon:2012uc}. More recent simulations have instead studied TDEs of main-sequence stars and included relativistic effects, which presumably play an important role for ``deep encounters" (\cite{Gafton:2019bzf}), where the star travels very close to the BH horizon.

However, TDE simulations are extremely challenging because they need to include physics at very different scales, which makes it hard to resolve stars and narrow debris streams. More precisely, let us consider a typical scenario in which a $1 M_{\odot}$ star is disrupted by a $10^6 M_{\odot}$ SMBH. The characteristic tidal radius \cite{Rees:1988bf}
\begin{equation}
    r_t \equiv R_\star \Bigg(\frac{M_{\rm BH}}{M_\star}\Bigg)^{1/3} ~,
    \label{eq:rtRees}
\end{equation}
is around $20 r_s$, where $r_s$ is the Schwarzschild radius of the BH and $R_\star$ is the radius of the star. This naturally leads to the separation of the physical scale
\begin{equation}
    r_t \gg r_s  > R_\star ~.
\end{equation}
The challenges in the simulation thus provide a strong motivation to come up with a simplified TDE model that is computationally cheap and capable of capturing the essential physical processes. In \cite{Lodato:2008fr,Kesden:2011ee,Kesden:2012qb,Servin:2016sog,Coughlin:2019pqk,Golightly:2019jib,Coughlin:2022ubg}, various approximation methods have been used to make predictions for TDE. For example, the star is assumed to be in hydrostatic equilibrium throughout and follow a nearly parabolic orbit until it reaches the pericenter (alternatively, the tidal radius). Then the star is disrupted instantaneously and each fluid element is modeled as a free particle, which follows its own ballistic orbit in the gravitational field of the black hole. This approximation is also known as frozen-in approximations (\cite{1982ApJ...262..120L,Lodato:2008fr}). This simplified model yields good agreement with simulations of the complete disruption of 5/3-polytropic (or equivalently, low-mass) stars.\footnote{As noted in \cite{Golightly:2019heg}, there are notable discrepancies between
the numerically obtained fallback rates, and those predicted by the frozen-in approximation, for higher mass and evolved stars.}


It is worth noting that the key missing ingredient in the above approximation is the tidal deformation of the star, which has a dynamical nature and arises from the excitation of the stellar oscillation modes. For this reason, the tidal bulge of the star is not aligned with the radial direction, as is observed in hydrodynamical simulations. This is critical to determining where the star is disrupted and the shape of the mass fallback rate.

\begin{figure*}
\centering
\includegraphics[width=0.85\textwidth]{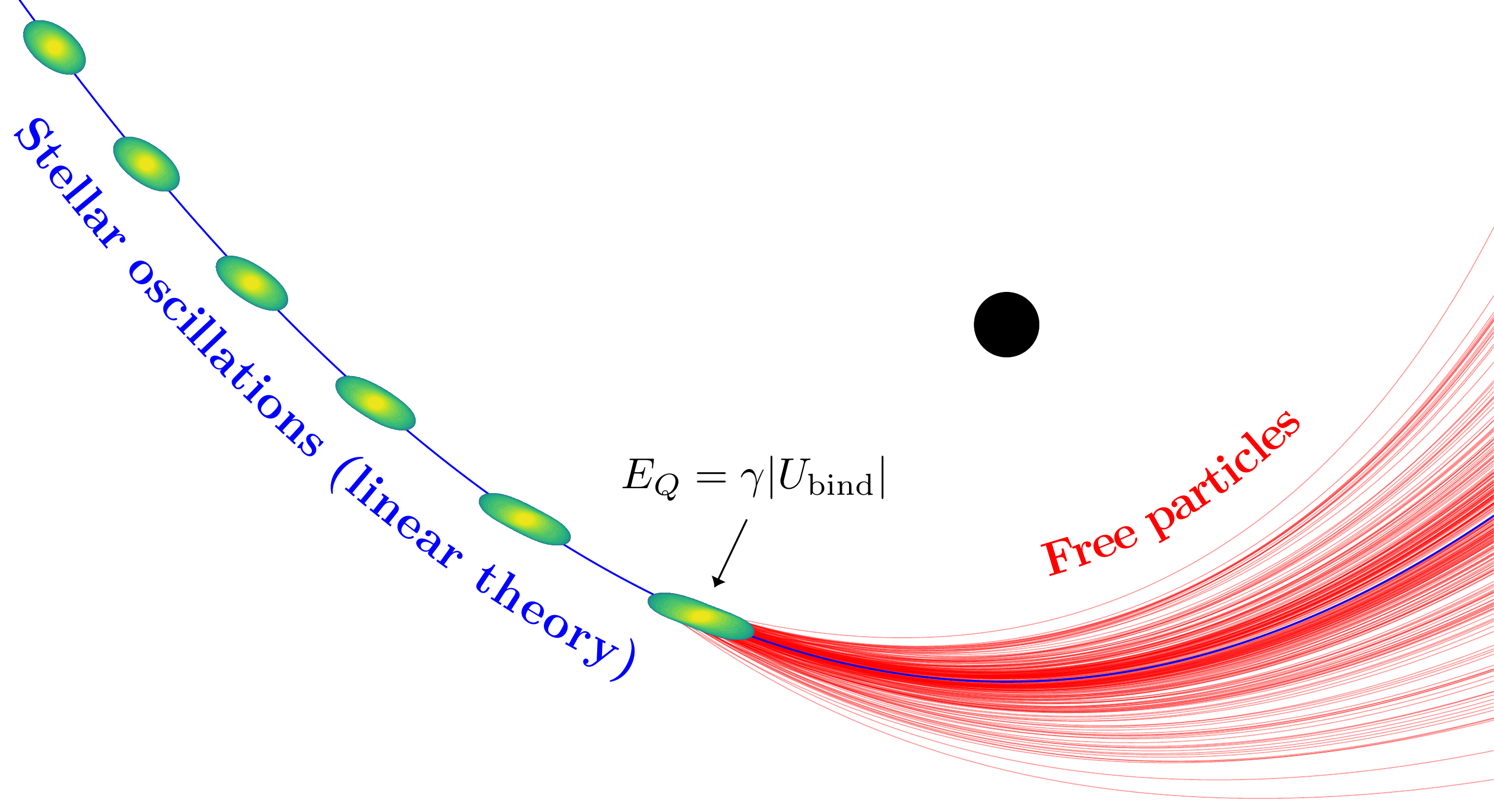}
\caption{Illustration of our two-stage model of TDEs. In the first stage, we compute the excitation of the stellar oscillation modes due to the tidal field of the BH. When the tidal energy $E_Q$ becomes a fraction $\gamma$ of the gravitational binding energy $U_{\rm bind}$, we transition into the second stage, where each fluid element is modeled as a free particle orbiting the BH. The value of $\gamma$ can be pre-calibrated by matching to numerical simulations. The figure depicts a tidal disruption of a $1M_\odot$ MESA MAMS star, on an orbit with $\beta\approx1.76\beta_{\rm crit}\approx4.44$ (corresponding to a pericenter distance $R_p=11GM_{\rm BH}$) by a BH with mass $M_{\rm BH}=10^6M_\odot$. All elements in the figure are drawn to scale, including the BH size, the density profile of the star and the randomly sampled free particle trajectories after the disruption.}
\label{fig:money-figure} 
\end{figure*}

In this paper, we build a simple two-stage model of TDEs, sketched in Fig.~\ref{fig:money-figure}, which incorporates the dynamical tidal effects. 
In the first stage, the star travels towards the black hole following quasi-parabolic orbits. We consider an effective description of the star as a point particle with multipole moments,
\begin{equation}
\label{eq:action}
    S = -\int \dd\tau  \Bigg(M_{\star} + \frac{1}{2}Q_{ij} E^{ij}  + \mathcal{L}_Q(Q_{ij},\dot Q_{ij}) + \ldots\Bigg)  ~,
\end{equation}
where $E_{ij}$ is the external electric parity part of the tidal field and $Q_{ij}$ is the induced quadrupole moment. The ellipsis represents higher-order multipoles suppressed by powers of $R_\star / r_t$. At the linear theory level, the dynamical structure of $Q_{ij}$ is captured by the linear stellar perturbation theory.\footnote{This method to include dynamical tides has been intensively studied in the context of asteroseismology (\cite{2010aste.book.....A,smeyers2011linear}), tidal excitations \cite{Schenk:2001zm,Weinberg:2011wf}, dynamical tidal capture (\cite{1977ApJ...213..183P}), etc.} In the Lagrangian picture, the stellar perturbation is equivalent to a collection of driven harmonic oscillators, while the BH tidal field can be viewed as the driving force that pumps a tidal energy
\begin{equation}
\label{eqn:E_Q}
    E_Q = - \frac{1}{2} \int_{-\infty}^\tau \dot Q_{ij}(\tau') E^{ij}(\tau') \dd\tau'~.
\end{equation}
As the star approaches disruption, the perturbations gradually become nonlinear. In our model, we extrapolate linear perturbations up to the onset of the disruption. We define this as the point where the tidal energy is comparable with the stellar binding energy
\begin{equation}
\label{eq:TDE_criteria}
    E_Q = \gamma |U_{\rm bind}| ~,
\end{equation}
where $\gamma$ is a calibration factor that can be pre-computed by fitting to the critical impact parameter
obtained in hydrodynamical simulations, as it only depends on the stellar structure. The deformed density profile can be evaluated by integrating the tidal response over the star's trajectory. When Eq.~\eqref{eq:TDE_criteria} is satisfied, the stellar perturbations are no longer stable, and the star is disrupted. In the second stage, we treat each fluid element inside the star as a free-falling particle, which allows us to evaluate the distribution of the orbital energy of the debris and the mass fallback rate. Even though most of the calculations in this model need to be done numerically, the whole computation only takes around one minute, much cheaper compared with the full hydrodynamical simulation.

The rest of the paper is structured as follows. In Section~\ref{sec:Model}, we illustrate our two-stage model in greater detail, including the stellar oscillation theory within external tidal fields, the tidal disruption criterion and the mass fallback rate. In Section~\ref{sec:Results}, we apply our model to the tidal disruption of the middle-age main-sequence (MAMS) with mass $0.15-8M_{\odot}$. We conclude in Section~\ref{sec:con} with a discussion of various approximations made in our work and potential future extensions. 

\section{A two-stage Model}
\label{sec:Model}

In this section, we give a thorough discussion of our two-stage model for TDE. In Section~\ref{subsec:Modelstage1}, we model the star as a point particle with multipole moments, which allows us to calculate the tidal deformation of the star. Then in Section~\ref{sec:TDE_criterion}, we discuss the TDE criterion in our model. Finally, after the star is disrupted, we model the debris fluid elements as free-falling particles to compute the mass fallback rate in Section~\ref{sec:stage-2}.

\subsection{Stage 1: stellar oscillations}
\label{subsec:Modelstage1}

In the first stage, we describe the motion of a star using multipole expansion methods, modeling the extended object as a point particle parameterized by its worldline proper time $\tau$ with mass monopole $M_\star$ and quadrupole moments $Q_{ij}(\tau)$ that encode the internal stellar dynamics (\cite{Goldberger:2004jt, Goldberger:2005cd}). It is straightforward to generalize the model to high-order multipoles. In this work, we focus on the leading quadrupole electric-type, i.e.\ parity even, quadrupole moment,\footnote{\cite{Binnington:2009bb} has pointed out that the magnetic multipole contributions is much smaller than the electric ones. Therefore, we do not include them in this paper.} which couples to the electric-type tidal field $E_{ij} = \partial_i \partial_j \Phi_{\rm BH}$, where $\Phi_{\rm BH}$ is the gravitational potential of the BH. Higher-order multipole moments are suppressed as they arise from higher derivatives of $\Phi_{\rm BH}$. 

The effective action of such model is given in Eq.~\eqref{eq:action}, where $\mathcal{L}_Q$ captures the internal stellar dynamics. We can write down the Euler-Lagrangian equation
\begin{equation}
    \frac{\dd}{\dd\tau}  \frac{\partial }{\partial \dot{Q}_{ij}}\mathcal{L}_Q - \frac{\partial \mathcal{L}_Q}{\partial Q_{ij}} = - \frac12 E^{ij} ~,
    \label{eq:EOMSLQ}
\end{equation}
which can be formally solved using the retarded Green's function method
\begin{flalign}
     Q_{ij}  (\tau) = -\frac{1}{2} \int_{\infty}^{\tau} \dd \tau' G^{}_{\rm ret} (\tau - \tau') E_{ij} (\tau') ~.
    \label{eq:expectvalQ}
\end{flalign}

As famously known from the linear response theory, the retarded Green's function can be decomposed into contributions from various eigenmodes of the system
\begin{flalign}
    G_{\rm ret} (\omega)  = - 2 \sum_n \frac{\lambda_n }{\omega^2_{n} - (\omega + i \epsilon)^2} ~, 
\label{eq:QexpectationQs}
\end{flalign}
with $\omega_n$ the eigenmode frequency and $\lambda_n$ the corresponding quadrupole overlap coefficient. Here, we have used the retarded $i\epsilon$-prescription to guarantee the causality of the Green's function (\cite{peskin2018introduction}).

The details of the response function can be extracted from the stellar fluid linear perturbation theory, with action (\cite{1964ApJ...139..664C,Schenk:2001zm,2003MNRAS.339...25R,Chakrabarti:2013xza})
\begin{equation}
\label{eq:fluid_lag}
    S = \int \dd^4 x \Bigg\{\frac{1}{2} \rho_0 \Big[\dot{\boldsymbol{\xi}}^2 - \boldsymbol{\xi} \cdot \mathcal{D} \boldsymbol{\xi}\Big]- \delta\rho \Phi_{\rm BH}\Bigg\} ~,
\end{equation}
where the Hermitian differential operator $\mathcal{D}$ acts on the fluid Lagrangian displacement $\boldsymbol{\xi}$ as 
\begin{equation}
    \mathcal{D} \boldsymbol{\xi} =-\nabla\biggl\{\left[\frac{c_s^2}{\rho_0}+4 \pi G \Delta^{-1}\right] \nabla \cdot\left(\rho_0 \boldsymbol{\xi}\right)\biggr\} ~,
\end{equation}
with $\rho_0(\boldsymbol{x})$ the background density profile and $c_s$ the local fluid speed of sound
\begin{equation}
    c_s^2(r) = \Big(\frac{\partial p}{\partial \rho}\Big)_{S} ~.
\end{equation}
The Eulerian density perturbation can be extracted from the continuity equation
\begin{equation}
\label{eq:fluid_eigen}
    \delta \rho = - \nabla \cdot (\rho_0 \boldsymbol{\xi}) ~.
\end{equation}
We then decompose the stellar fluid displacement into 
time-dependent amplitudes and spatial eigenfunctions,

\begin{flalign}
    \boldsymbol{\xi} (t,\boldsymbol{x})  = \sum_{n \ell m} q_{n \ell m}(t)  \boldsymbol{\xi}_{n \ell m } (\boldsymbol{x}) ~,
    \label{eqn:xi}
\end{flalign}
with $n$, $\ell$, $m$ the principal, angular and azimuthal quantum numbers. In the vector spherical harmonic basis, we can conveniently decompose the eigenfunctions as
\begin{flalign}
\begin{split}
\boldsymbol{\xi}_{n \ell m} (\boldsymbol{x})&= \xi_{n \ell}^{\mathrm{R}}(r) \boldsymbol{Y}_{ \ell m}^{\rm R}(\theta,\phi)+\xi_{n \ell}^{\mathrm{E}}(r) \boldsymbol{Y}_{ \ell m}^{\rm E}(\theta,\phi)~,  
\end{split}
\end{flalign}
with the vectors $\boldsymbol{Y}_{\ell m}^{\rm R}= \hat{\boldsymbol{r}} Y_{\ell m}$  and $\boldsymbol{Y}_{ \ell m}^E = r \nabla Y_{\ell m}$ the radial and poloidal components. Furthermore, the fluid perturbation eigenfunctions are orthogonal with respect to the inner product,
\begin{flalign}
    \int \dd^3 x\, \rho_0  \boldsymbol{\xi}_{n'\ell' m'} \boldsymbol{\xi}_{n\ell m} = M_\star R_{\star}^2 \mathcal{N}_{n\ell } \delta_{n' n} \delta_{\ell' \ell} \delta_{m m'}~, 
    \label{eq:orthogonalityrel}
\end{flalign}
with dimensionless normalization constant $\mathcal{N}_{n \ell}$ 
\begin{flalign}    
\begin{split}    
\mathcal{N}_{n \ell} 
=& \frac{1}{M_\star R_\star^2}\int \dd r\,  \rho_{0}\Big( (r \xi_{n \ell}^{\mathrm{R}})^2+(\ell+1) (r \xi_{n \ell}^{\rm E})^2   \Big) ~.
\label{eq:Normalizationconstant}
\end{split}
\end{flalign}

Plugging the above decomposition into the Lagrangian in Eq.~\eqref{eq:fluid_lag}, we find that it simply becomes a collection of driven harmonic oscillators with eigenfrequencies $\omega_{n\ell}$,
\begin{flalign}    
\mathcal{L}  =&   \sum_{n \ell m} \left( \frac{M_\star R_\star^2 \mathcal{N}_{ n \ell }}{2} \left( \dot{q}_{n \ell m}^2 - \omega_{n \ell }^2  q_{n \ell m}^2  \right) +  q_{n \ell m } F_{n \ell m}   \right) ~,
\label{eq:hsoq}
\end{flalign}
with driving force 
\begin{equation}
    F_{n \ell m} (t) = - \int \dd^3 x\, \delta \rho_{n \ell  m} (t,\boldsymbol{x}) \Phi_{\rm BH}(t,\boldsymbol{x})~.
\end{equation}

For the quadrupole perturbation $\ell=2$, it is convenient to introduce the tensorial spherical harmonics $ \mathcal{Y}_{\ell m}^{ij}$ defined as
\begin{equation}
    Y_{\ell m}  \equiv \mathcal{Y}_{\ell m}^{ij} \Big(\hat{r}_i \hat{r}_j - \frac{1}{3} \delta_{ij}\Big) ~.
\end{equation}
In the frequency domain, the solution takes the form
\begin{flalign}
\label{eq:fluid_amp_sol}
    q_{n 2 m} (\omega) = \frac{1}{\omega_{n}^2 - (\omega+i\epsilon)^2} \frac{F_{n 2 m} (\omega)}{\mathcal{N}_{n 2}} ~,
\end{flalign}
where we defined $\omega_{n}\equiv\omega_{n2}$ for short. The driving force in this basis is given by
\begin{flalign}
\begin{split}    
     F_{n 2 m} (t) 
     =&  \sqrt{\frac{8 \pi}{15}}  M_\star R_\star^{2} \mathcal{I}_{n 2}    \mathcal{Y}_{2 m}^{ij}  E_{ij} (t) ~,
\end{split}
\end{flalign}
with $\mathcal{I}_{n \ell}$  the dimensionless fluid mode overlap integral
\begin{flalign}
\label{eq:overlap_int}
\mathcal{I}_{n \ell}=  \frac{1}{M_\star R_\star^{\ell}}  \int \dd r\, r^{\ell+1} \rho_{0}\Big(\xi_{n \ell}^{\mathrm{R}}+ (\ell+1) \xi_{n \ell}^{\mathrm{E}}\Big) ~.
\end{flalign}
The fluid amplitude solution Eq.~\eqref{eq:fluid_amp_sol} along with the eigenfunction in Eq.~\eqref{eq:fluid_eigen} completely determine the linear stellar fluid motion within the external tidal field.

Matching to our effective point particle description, the induced quadrupole moment due to fluid perturbations can be expressed as
\begin{flalign}
\begin{split}
Q^{ij}_{} 
&= \int \dd^3 x\, r^2 \; \delta \rho (t,\boldsymbol{x})\; \Big(\hat{r}^i \hat{r}^j - \frac{1}{3}\delta^{ij}\Big) \\
&= \sum_{n } \sqrt{\frac{32 \pi}{15}}  M_\star R_\star^2 \mathcal{I}_{n 2}   q_{n 2 m}  \mathcal{Y}_{2 m}^{ij} ~,
\end{split}
\label{eq:inducedmultip}
\end{flalign}
which allows us to compute the quadrupole overlap coefficient in Eq.~\eqref{eq:QexpectationQs}
\begin{flalign}
    \lambda_{n} = \frac{16 \pi}{15} M_\star R_\star^2  \frac{\mathcal{I}_{n 2}^2}{\mathcal{N}_{n2}}~.
    \label{eq:tidalparamfluid-dimq}
\end{flalign}

The stellar fluid oscillation modes are classified according to their restoring force and dispersion relation (\cite{1941MNRAS.101..367C, smeyers2011linear}). For
acoustic waves, also known as pressure modes ($p$-modes), pressure serves as the restoring force
and leads to the dispersion relation
\begin{equation}
    \omega^2 \sim c_s^2 |\boldsymbol{k}|^2 ~,
\end{equation}
where we denote the wavenumber $\boldsymbol{k} = k_r \hat{\boldsymbol{r}} + \boldsymbol{k}_{\rm H}$. Acoustic waves have high oscillation frequencies and are dominated by the radial fluid displacement. They characterize the oscillations in the outer layers of the star. The second type of waves are gravity waves ($g$-modes), which have low oscillation frequencies with dispersion relation
\begin{equation}
    \omega^2 \sim N^2 \frac{\left|\boldsymbol{k}_{\mathrm{H}}\right|^2}{|\boldsymbol{k}|^2} ~.
\end{equation}
The characteristic frequency for $g$-modes is the Brunt-V\"{a}is\"{a}l\"{a} frequency 
\begin{equation}
    N^2 \equiv g^2\Bigg(\frac{1}{c_{\rm eq}^2}  - \frac{1}{c_s^2} \Bigg) ~.
\end{equation}
with the local gravitational acceleration function $g(r)\equiv G M_\star(r)/r^2$ and the local equilibrium sound speed 
\begin{equation}
    c_{\rm eq}^2(r) \equiv \frac{p_0'(r)}{\rho_0'(r)} ~.
\end{equation}
The gravity waves are dominated by the horizontal displacement and oscillate deep in the stellar interior. The fluid fundamental mode ($f$-mode) sits between the gravity modes and the pressure modes. It corresponds to the surface gravity wave whose restoring force primarily comes from gravity at the stellar surface. Both the radial and horizontal displacements of the $f$-mode peak at the surface and decay quickly inward. All the $p$- and $g$- modes have infinite number of overtones, $n=1,2,\ldots$ for $p$-modes and $n=-1,-2,\ldots$ for $g$-modes. The absolute value $|n|$ also represents the number of nodes in the radial eigenfunction. The $f$-mode is denoted by $n=0$ and has zero nodes.

In practice, once we know all the eigenfrequencies and eigenfunctions of fluid oscillations for a given star, we are able to use Eqs.~\eqref{eq:Normalizationconstant},~\eqref{eq:fluid_amp_sol},~\eqref{eq:overlap_int} and \eqref{eq:inducedmultip} to compute the induced quadrupole moment of the star.

\subsection{Tidal energy and disruption criterion}
\label{sec:TDE_criterion}

The model of stellar oscillations described above holds as long as the perturbations are small. When they become large enough, the star is disrupted. We define a disruption criterion by quantifying the energy stored in the stellar degrees of freedom due to tidal interactions. We can gain physical insight by looking at a simpler system such as a particle in a potential well with an external driving force. As long as the energy of the particle is smaller than the binding energy of the well, the particle remains confined. When it gains more energy from the driving force than the binding energy, the particle escapes.

With this example in mind, let us calculate the tidal energy pumped into the star. This is given by
\begin{flalign}
    E_Q  =  \dot{Q}_{ij} \frac{\partial\mathcal{L}_Q}{\partial\dot{Q}_{ij}}- \mathcal{L}_Q ~.
\end{flalign}
Using the equations of motion in Eq.~\eqref{eq:EOMSLQ}, we obtain
\begin{equation}
    \frac{\dd}{\dd\tau} E_Q = -\frac12\dot Q_{ij} E^{ij} ~,
\end{equation}
which then leads to the formula in Eq.~\eqref{eqn:E_Q}.

The depth of the potential well is given by the gravitational binding energy of the star $U_{\rm bind}$, where
\begin{equation}
    U_{\rm bind} = - \int_0^{M_\star} \frac{G m(r) dm}{r} ~, \quad   m(r) \equiv \int_0^r \rho_0 dV ~.
\end{equation}
We thus expect the star to be disrupted when $E_{Q} \sim |U_{\rm bind}|$. Because we extrapolate our linear oscillations model until the star is disrupted, we allow for a $\mathcal O(1)$ correction to the above relation. We thus introduce a calibration factor $\gamma$ and define the tidal disruption criterion as
\begin{equation}
    E_{Q} / |U_{\rm bind}| \equiv \gamma \simeq \mathcal{O}(1)~. 
    \label{eq:disruptioncriteria}
\end{equation}

We determine $\gamma$ by matching to the results of hydrodynamical simulations as follows. The depth of a TDE is often quantified by the parameter
\begin{equation}
    \beta \equiv \frac{r_t}{r_p} ~,
\end{equation}
where $r_t$ is the characteristic tidal radius given in Eq.~\eqref{eq:rtRees} and $r_p$ is the pericenter distance. Hydrodynamical simulations found that the critical value of $\beta$, above which the star is fully disrupted, can be approximated as (\cite{Law-Smith:2020zkq, Ryu:2020huz})
\begin{equation}
    \beta_{\rm crit} \approx 0.5 \Big(\frac{\rho_c}{\bar \rho}\Big)^{1/3} ~.
    \label{eqn:beta-crit}
\end{equation}
where $\rho_c$ and $\bar\rho$ are the central and average densities of the star.\footnote{This formula can also be understood as the characteristic tidal radius of the stellar core.} We thus compute $\gamma$ in our model as
\begin{equation}
    \gamma = \max_\tau \{E_Q(\tau) / |U_{\rm bind}| \} ~,
    \label{eqn:gamma-def}
\end{equation}
on an orbit with $\beta=\beta_{\rm crit}$. Once the calibration factor is fixed for a certain stellar type, we can carry out computations for the same type of star on all kinds of orbits without having to re-compute $\gamma$.

\begin{figure}
\centering
\includegraphics{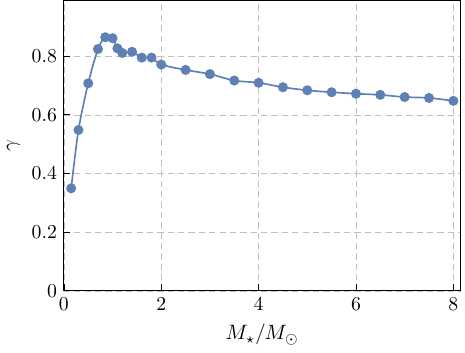}
\caption{Calibration factor $\gamma$, as defined in Eq.~\eqref{eq:disruptioncriteria}, as a function of the stellar mass for MESA stars. The calibration factor is computed as in Eq.~\eqref{eqn:gamma-def}, for an impact parameter equal to $\beta_{\rm crit}$, given in Eq.~\eqref{eqn:beta-crit}.}
\label{fig:gamma} 
\end{figure}

In Fig.~\ref{fig:gamma} we show the value of $\gamma$ for MESA MAMS stars. We see that $\gamma$ is indeed $\mathcal O(1)$, as expected from our potential well analogy. Its value increases with $M_\star$ for $M_\star<1M_\odot$, corresponding to convective stars. Conversely, it slightly decreases with $M_\star$ for $M_\star>1M_\odot$, corresponding to radiative stars.

\subsection{Stage 2: free particles}
\label{sec:stage-2}

In the second stage of the TDE, the star is already disrupted and the material inside the star is no longer bounded by the stellar gravitational potential. To the zeroth order approximation, we can treat each fluid element inside the star as a free-falling particle. More concretely, the full evolution of the second stage is controlled by the collisionless Boltzmann equation 
\begin{equation}
\label{eq:Boltzmann equation}
    \frac{\partial f}{\partial t} + \boldsymbol{v} \cdot \nabla_{\boldsymbol{x}} f + \boldsymbol{a} \cdot \nabla_{\boldsymbol{v}} f = 0 ~,
\end{equation}
where $f$ is the distribution function in phase space. For each fluid element, the acceleration is determined by
\begin{equation}
    \boldsymbol{a} = - \nabla_{\boldsymbol{x}} (\Phi_{\star} + \Phi_{\rm BH})
\end{equation}
where $p$ is the fluid pressure, $\rho$ is the density, $\Phi_\star$ is the stellar debris self-gravity 
and $\Phi_{\rm BH}$ is the gravitational potential of the BH. To capture the essential physics for the material fallback towards to the BH, we can ignore the pressure and self-gravity and write
\begin{equation}
    \boldsymbol{a} \simeq - \nabla_{\boldsymbol{x}} \Phi_{\rm BH} ~.
\end{equation}
Any function $f(E, L_z)$ of the orbital energy $E$ and angular momentum $L_z$ is a solution to Eq.~\eqref{eq:Boltzmann equation} because of the conservation laws for Newtonian orbits. In reality, it is also crucial to notice that ignoring pressure and self-gravity may lead to both qualitative and quantitative differences in the energy
distribution and the fallback rate compared to numerical simulations (\cite{2023MNRAS.526.2323F}).

In practice, we calculate the Lagrangian displacement at the disruption time $t_{\rm TDE}$ (i.e., when $E_Q=\gamma|U_{\rm bind}|$), which also allows us to obtain the density profile at that instant. After that, we sample the stellar fluid elements to determine the orbital energy distribution $\dd M/ \dd E$. In Newtonian gravity, we can further use the relation 
\begin{equation}
    E = -\frac{1}{2} \Big(\frac{2\pi G M_{\rm BH}}{T}\Big)^{2/3} 
\end{equation}
to calculate the mass fallback rate $\dd M / \dd T$, which, to leading order, is responsible for determining the observed lightcurve in a TDE (\cite{Coughlin:2022ubg}).

\section{Results}
\label{sec:Results}

\subsection{MESA stars on parabolic orbits}

The model described in Section~\ref{sec:Model} can be applied to any given orbit around the BH, and to any stellar model whose oscillation eigenfrequencies $\omega_n$ and eigenfunctions $\xi_n$ are known. We consider here MESA (\cite{2011ApJS..192....3P}) MAMS stars, with masses $M_\star\in[0.15M_\odot,8M_\odot]$, and compute their oscillation properties with GYRE (\cite{2013MNRAS.435.3406T}).\footnote{We define MAMS as having a hydrogen fraction of $0.35$. However, low-mass stars take much longer than the age of the Universe to become MAMS, so they are more likely to be well described as zero-age main sequence (ZAMS) stars, with a hydrogen fraction of $0.7$. For simplicity, in this paper we ignore this issue and assume all stars are MAMS.} For each star, we include modes with frequency $\omega_n\in[0.1\omega_\star,20\omega_\star]$, where $\omega_\star=\sqrt{GM_\star/R_\star^3}$ is the typical oscillation frequency of the star. The addition of modes outside this frequency range does not significantly change the results.

For simplicity, we consider here Newtonian parabolic orbits, though the model can in principle also be applied to general relativistic orbits in the Kerr spacetime. In cartesian coordinates, such that the pericenter is at $(r_p,0,0)$ and the orbit is within in the $xy$ plane, the nonzero components of the tidal field are given by
\begin{align}
E_{xx}&=(1-3\cos^2\phi)GM_{\rm BH}/r^3~,\\
E_{xy}&=3\sin\phi\cos\phi\, GM_{\rm BH}/r^3~,\\
E_{yy}&=(1-3\sin^2\phi)GM_{\rm BH}/r^3~,\\
E_{zz}&=GM_{\rm BH}/r^3~,
\end{align}
where $r$ is the radial distance and $\phi$ is the polar angle. From this, the quadrupole moment can be computed as in Eq.~\eqref{eq:inducedmultip}.

\begin{figure*}
\centering
\includegraphics{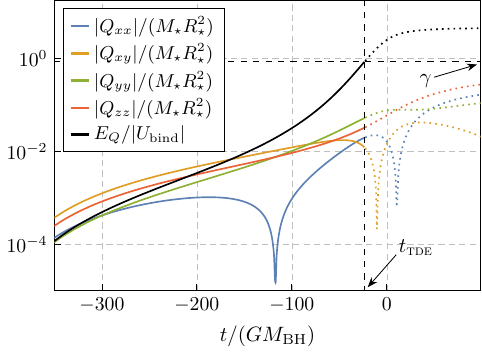}\qquad\qquad
\includegraphics{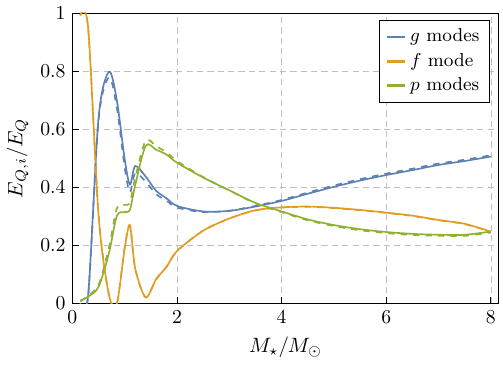}
\caption{\emph{Left:} time evolution of the components of the quadruple moment $Q_{ij}$ and the tidal energy $E_Q$, defined in Eq.~\eqref{eqn:E_Q}. The pericenter is at $t=0$ and the orbital parameters, as well as the star's and BH's masses, are the same as in Fig.~\ref{fig:money-figure}. We highlight the point where $E_Q/|U_{\rm bind}|=\gamma$, which corresponds to the disruption time $t_{\rm TDE}$. The first stage of our model ends at $t=t_{\rm TDE}$. The perturbative calculation at $t>t_{\rm TDE}$ is no longer applicable, therefore we draw all lines as dotted. \emph{Right:} fractional contribution (as defined in Eq.~\eqref{eqn:E_Q-decomposed}) of $g$-, $f$- and $p$-modes to the total tidal energy $E_Q$ at $t=t_{\rm TDE}$, as a function of $M_\star$. The solid lines assume $\beta=1.76\beta_{\rm crit}$, while the dashed line assume $r_p=17GM_{\rm BH}$. It is apparent that the result is largely insensitive to the pericenter distance.}
\label{fig:Q_E} 
\end{figure*}

\begin{figure*}
\centering
\includegraphics{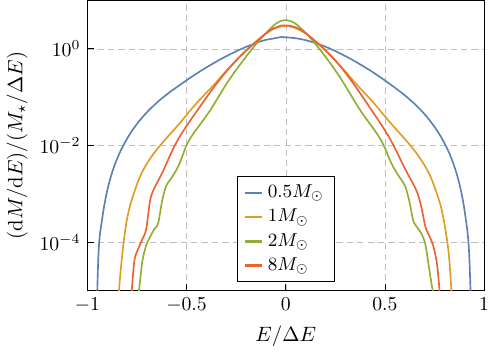}
\qquad\qquad
\includegraphics{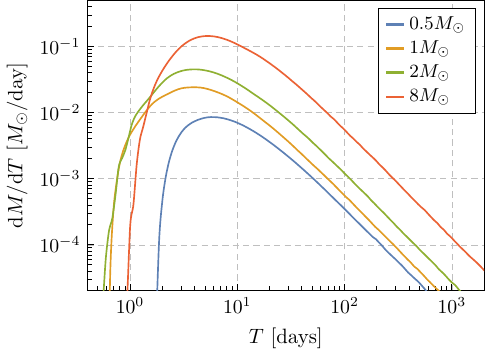}
\caption{Specific energy distribution (\emph{left panel}) and mass fallback rate (\emph{right panel}) for MESA MAMS stars with masses $0.5M_\odot$, $1M_\odot$, $2M_\odot$, and $8M_\odot$. All cases assume $M_{\rm BH}=10^6M_\odot$ and $\beta=1.76\beta_{\rm crit}$, which equals $2.61$, $4.44$, $5.60$, $4.37$ in the four cases respectively. The left panel is normalized according to Eq.~\eqref{eqn:deltaEdeltaT}, while the right panel displays physical units.}
\label{fig:dMdE-dMdT} 
\end{figure*}

We show in the left panel of Fig.~\ref{fig:Q_E} an example of the time evolution of the quadrupole moment $Q_{ij}$ and the tidal energy $E_Q$, with $\beta\approx1.76\beta_{\rm crit}$. In this example, the tidal energy increases monotonically and reaches the critical value $\gamma|U_{\rm bind}|$ before pericenter. Higher values of $\beta$ lead to an earlier disruption time. Conversely, for values of $\beta$ marginally above $\beta_{\rm crit}$, the disruption point can be located well after pericenter.

Each stellar oscillation mode contributes a term $E_{Q,n}$ to the tidal energy, which can be decomposed as
\begin{equation}
    E_Q=\sum_nE_{Q,n}=E_{Q,g}+E_{Q,f}+E_{Q,p}~.
    \label{eqn:E_Q-decomposed}
\end{equation}
In this expression, $E_{Q,g}$ includes all contributions from $g$-modes ($n<0$), $E_{Q,p}$ includes those from $p$-modes ($n>0$) and $E_{Q,f}=E_{Q,0}$ corresponds to the $f$-mode. In the right panel of Fig.~\ref{fig:Q_E}, we show the relative contributions of these terms to the tidal energy at the disruption time $t=t_{\rm TDE}$. We see that the way the energy is distributed in the various modes depends very sensitively on the stellar structure. $g$-modes dominate in stars with mass between $0.5M_\odot$ and $1M_\odot$, as well as above $4M_\odot$. $p$-modes contribute most significantly between $1M_\odot$ and $4M_\odot$, while the $f$-mode is dominant below $0.5M_\odot$. Remarkably, large changes in the pericenter distance $r_p$ do not significantly affect these conclusions. This suggests that the fractional energy contribution of $g$-, $f$- and $p$-modes in a given star is almost independent of the orbital parameters.

\subsection{Energy distribution, fallback rate, and density profile}

\begin{figure*}
\centering
\includegraphics[width=0.225\textwidth]{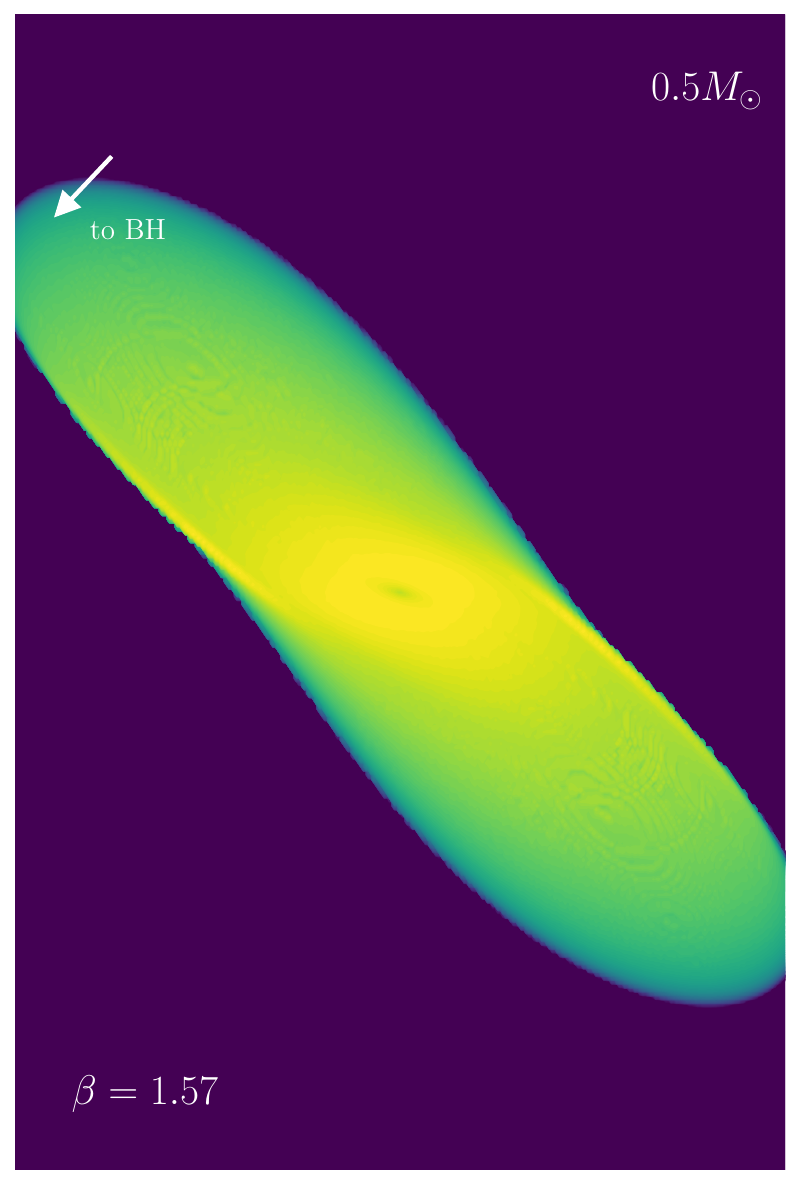}
\includegraphics[width=0.225\textwidth]{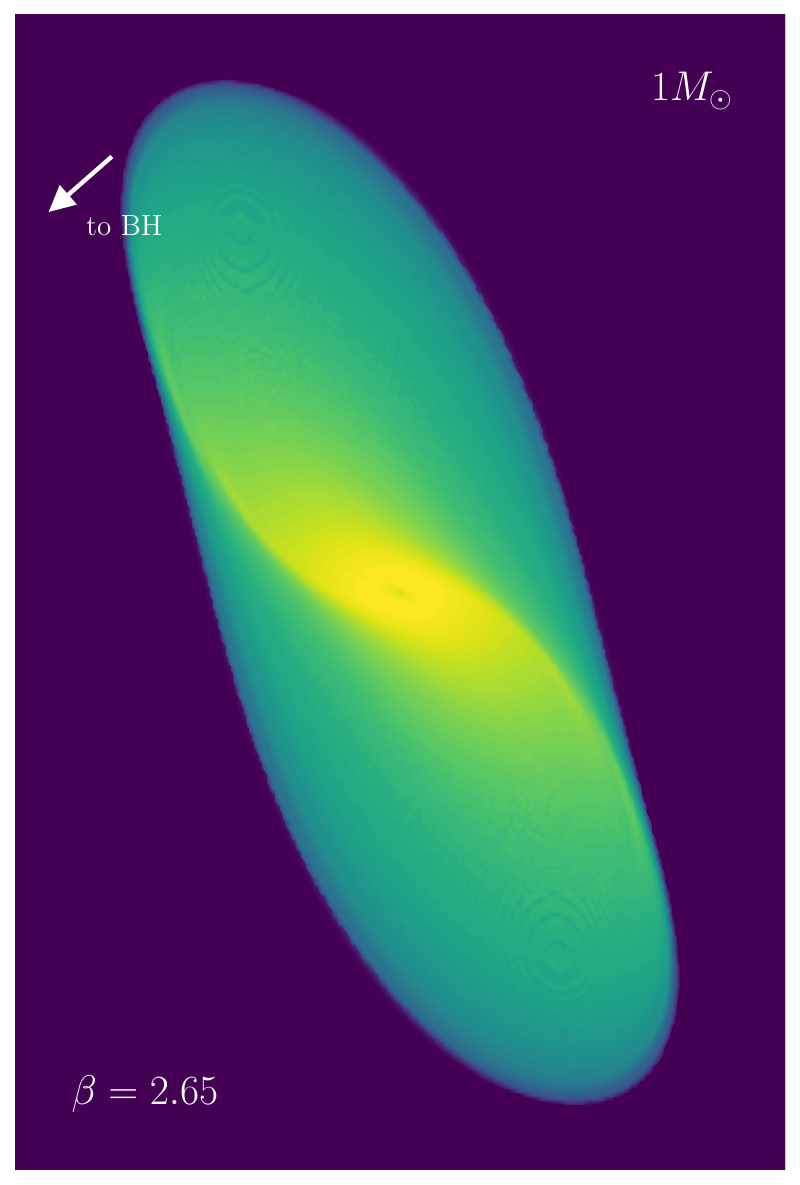}
\includegraphics[width=0.225\textwidth]{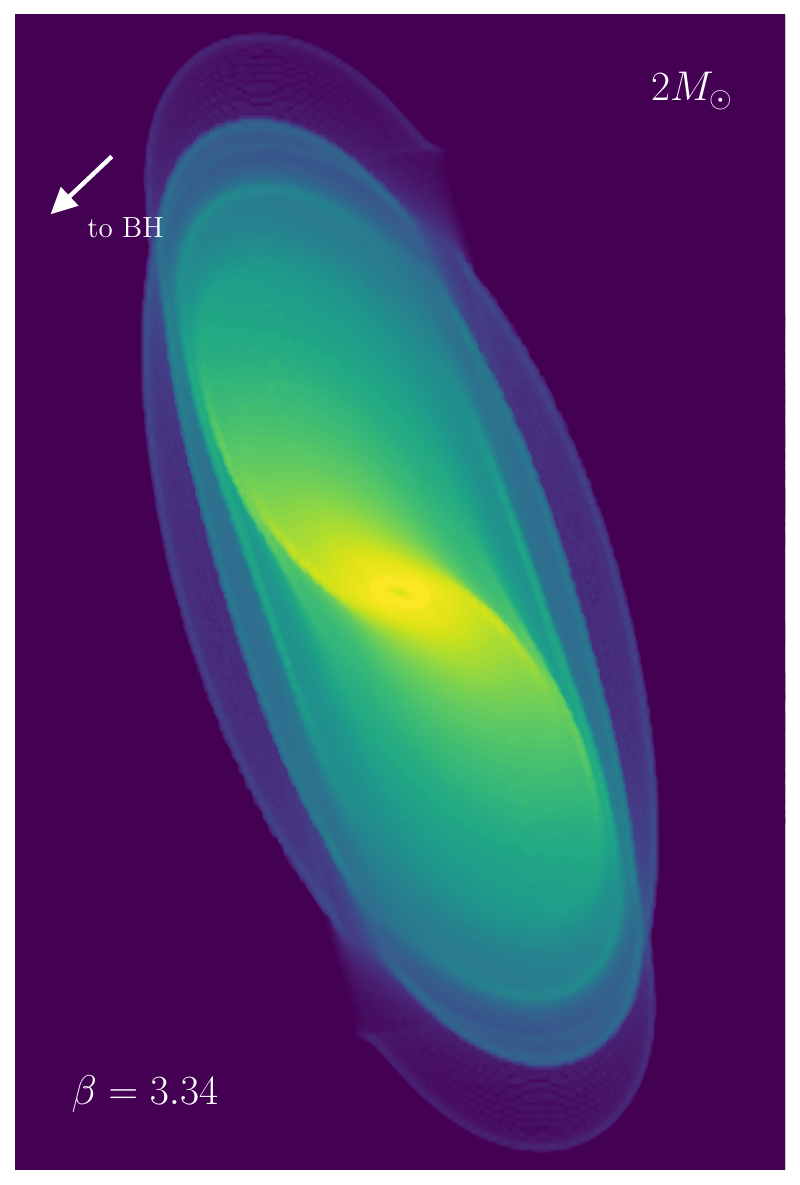}
\raisebox{-3.3pt}{\includegraphics[width=0.294\textwidth]{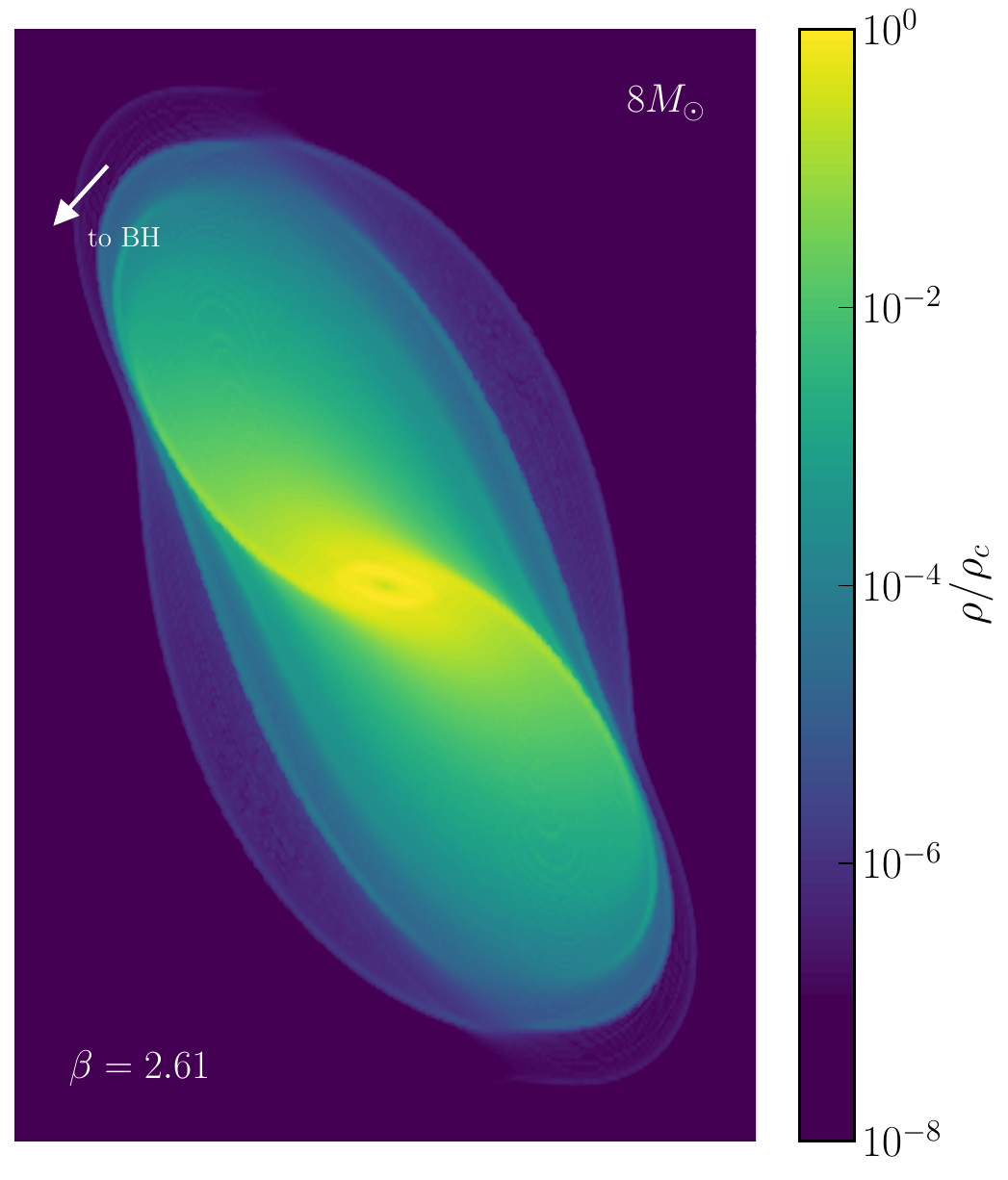}}
\includegraphics[width=0.225\textwidth]{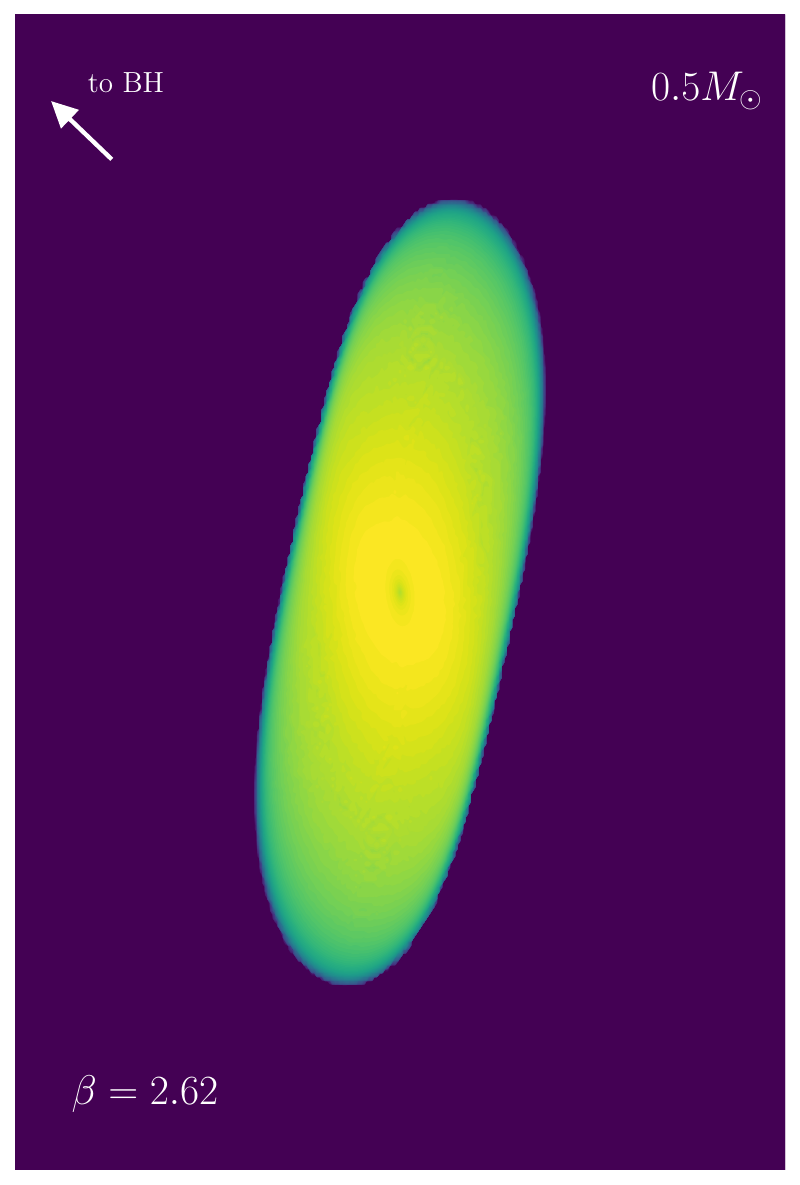}
\includegraphics[width=0.225\textwidth]{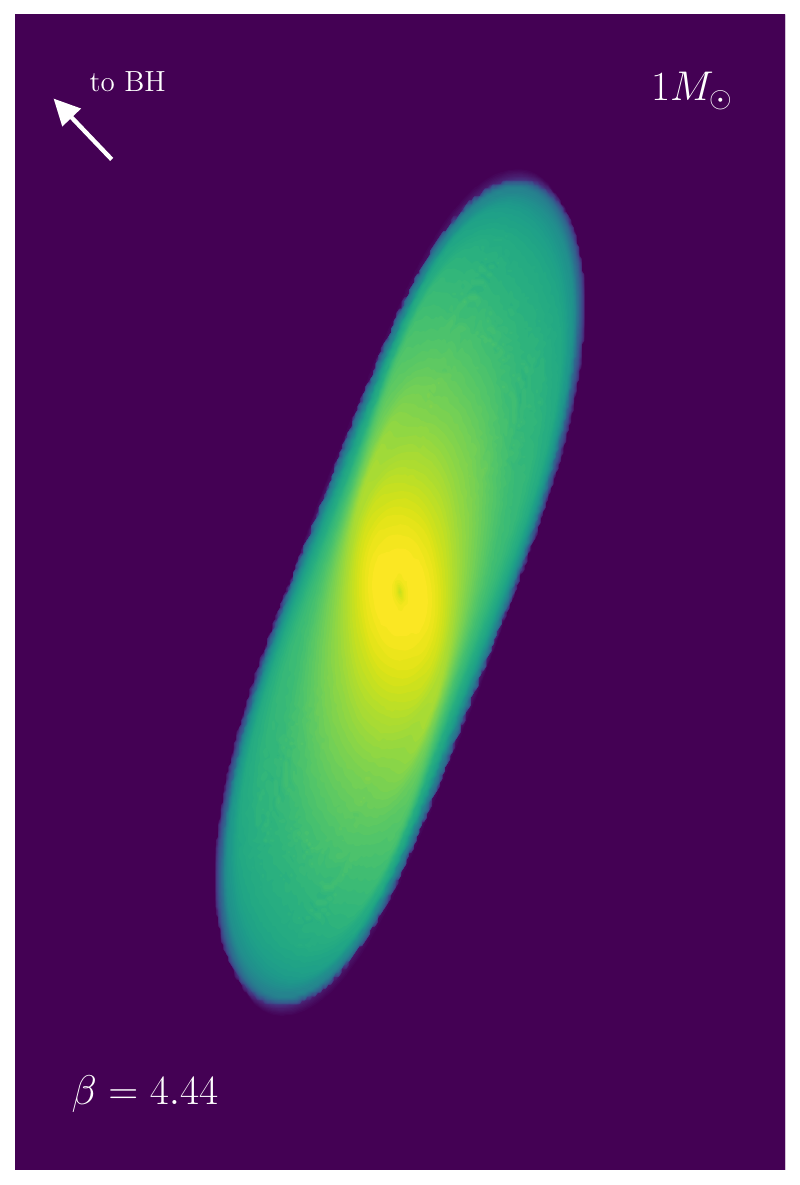}
\includegraphics[width=0.225\textwidth]{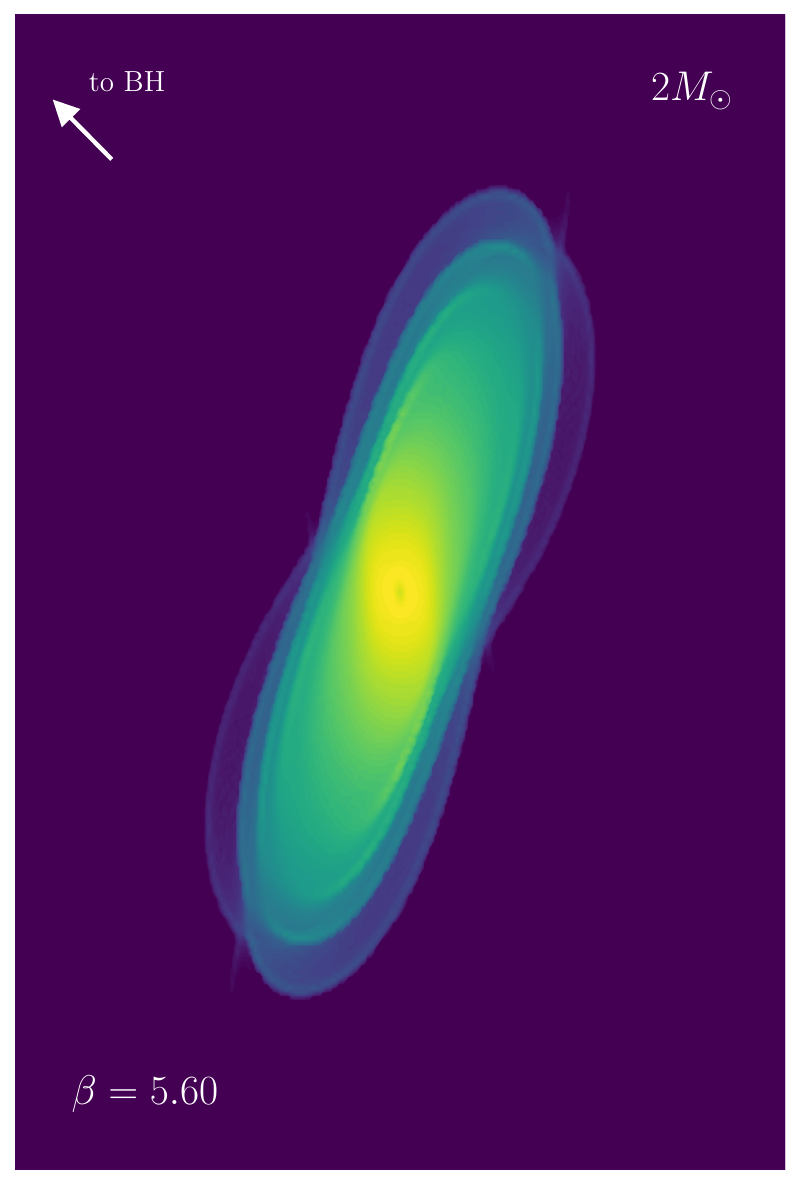}
\raisebox{-3.3pt}{\includegraphics[width=0.294\textwidth]{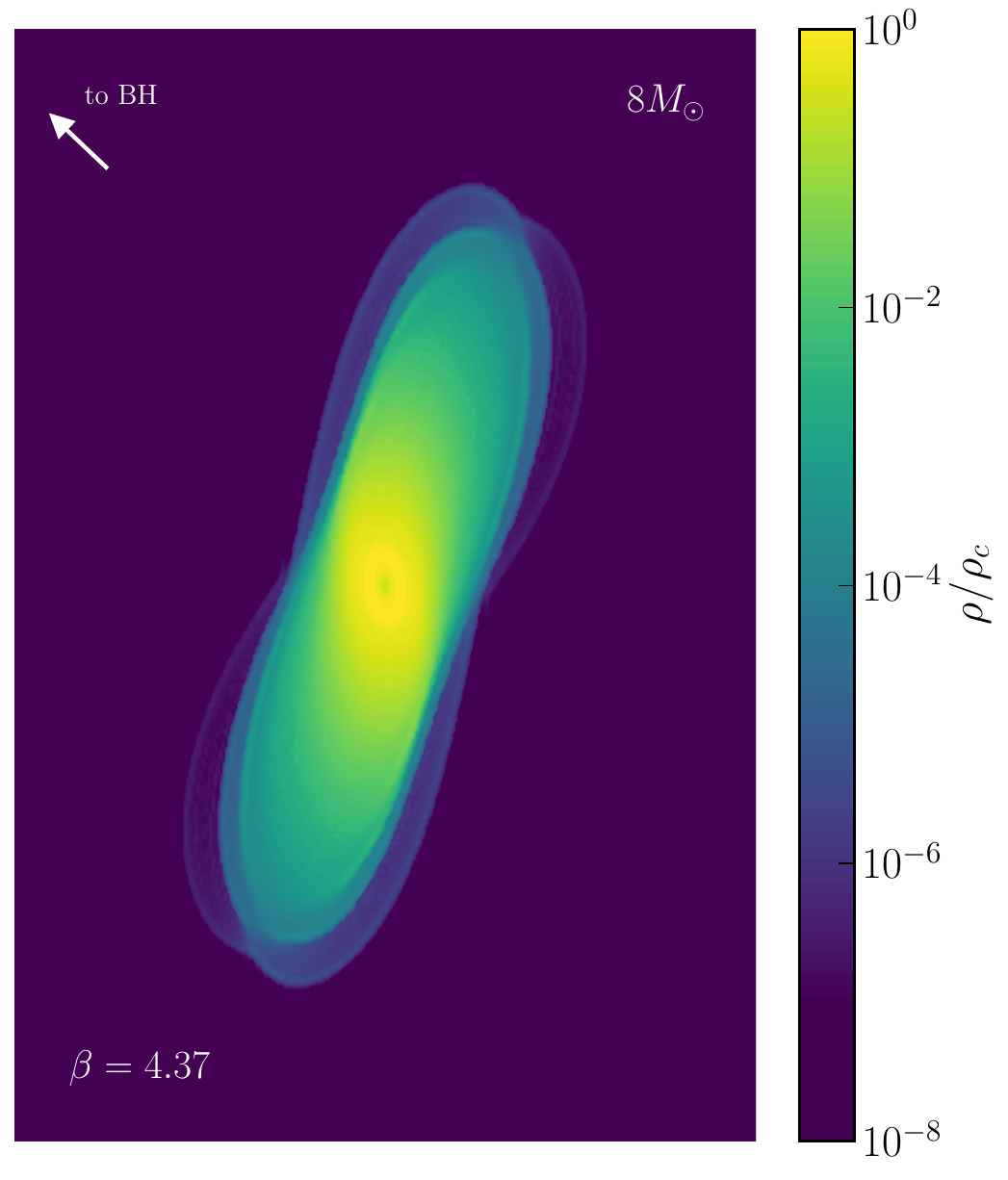}}
\includegraphics[width=0.225\textwidth]{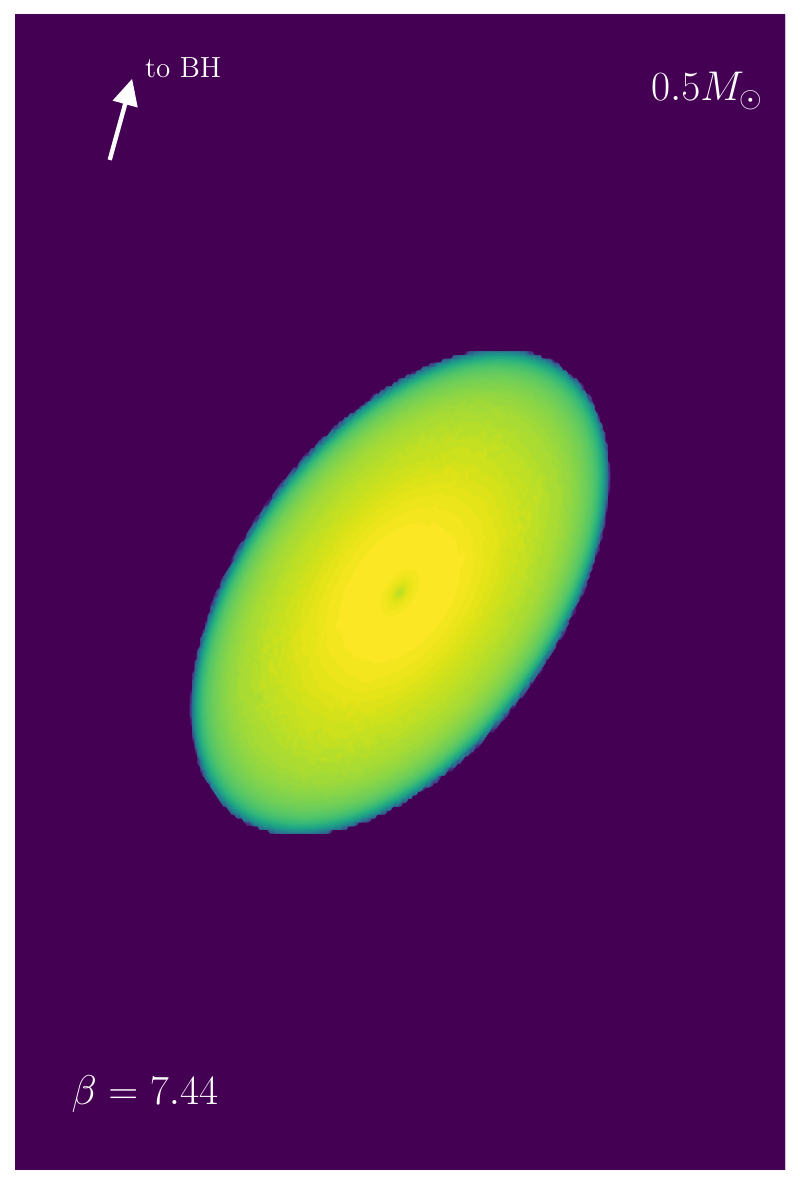}
\includegraphics[width=0.225\textwidth]{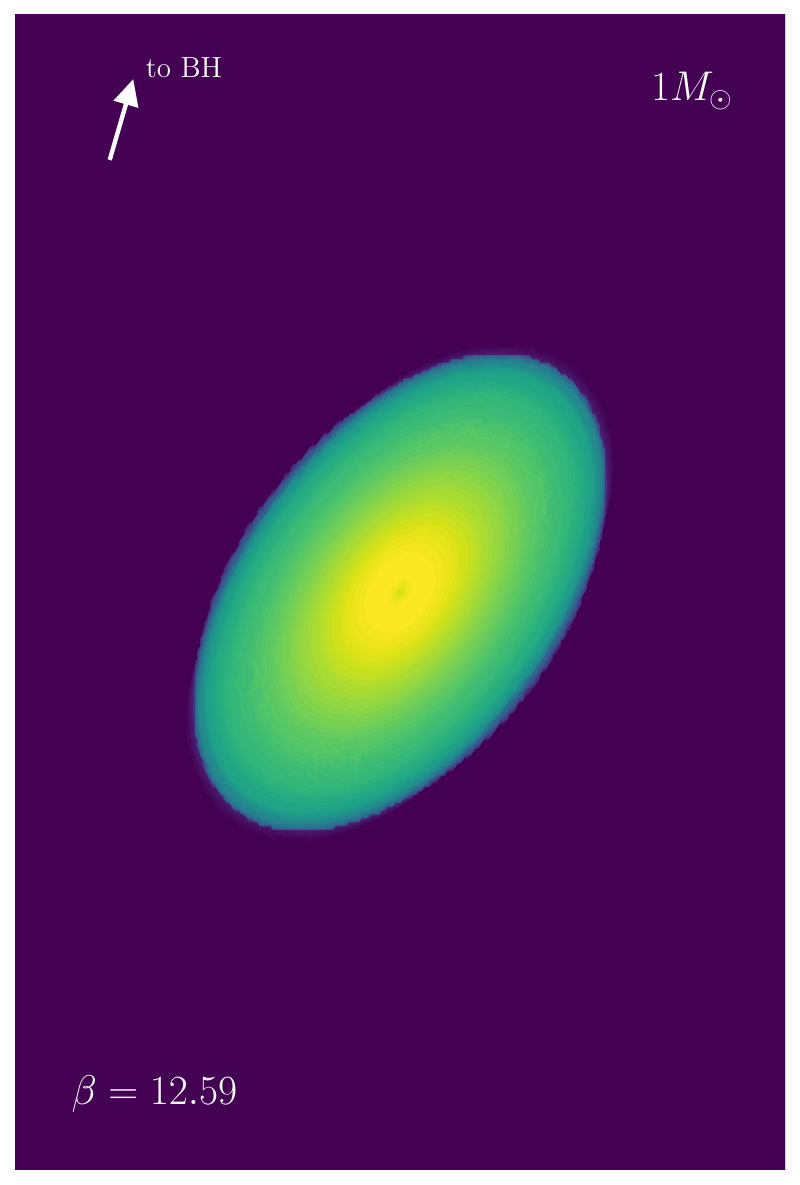}
\includegraphics[width=0.225\textwidth]{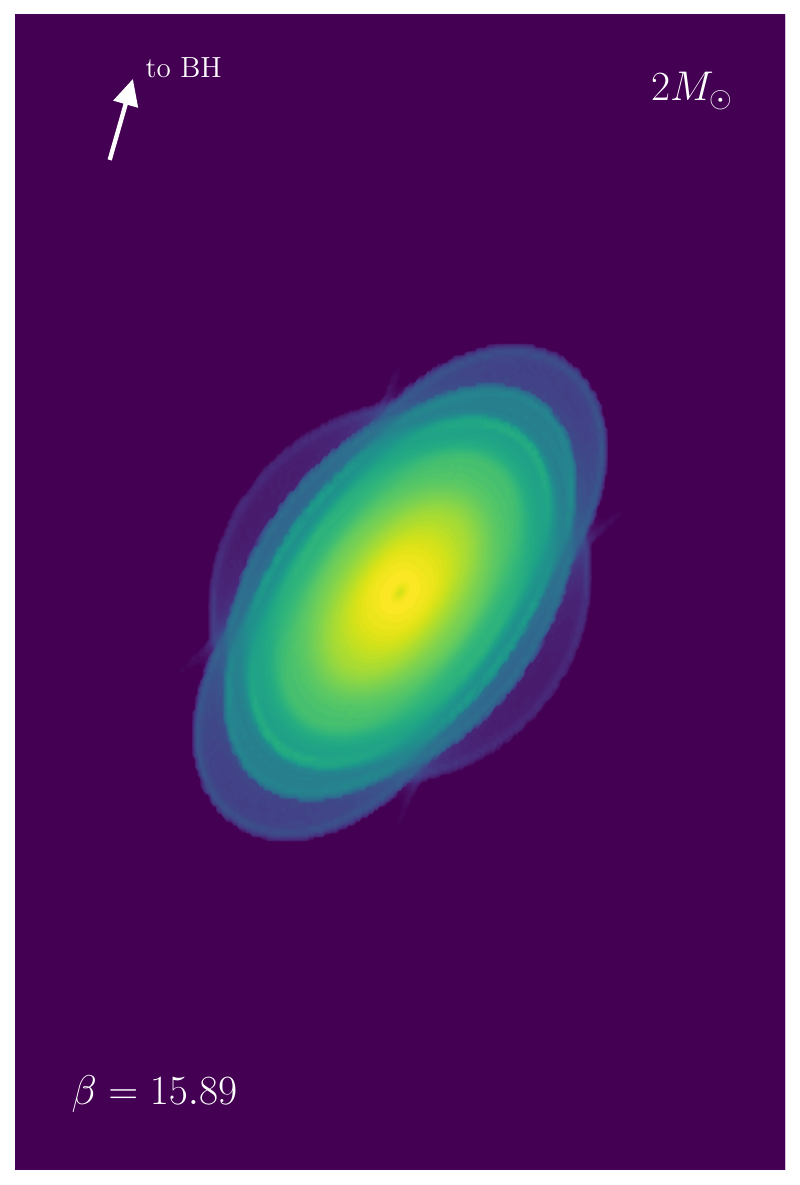}
\raisebox{-3.3pt}{\includegraphics[width=0.294\textwidth]{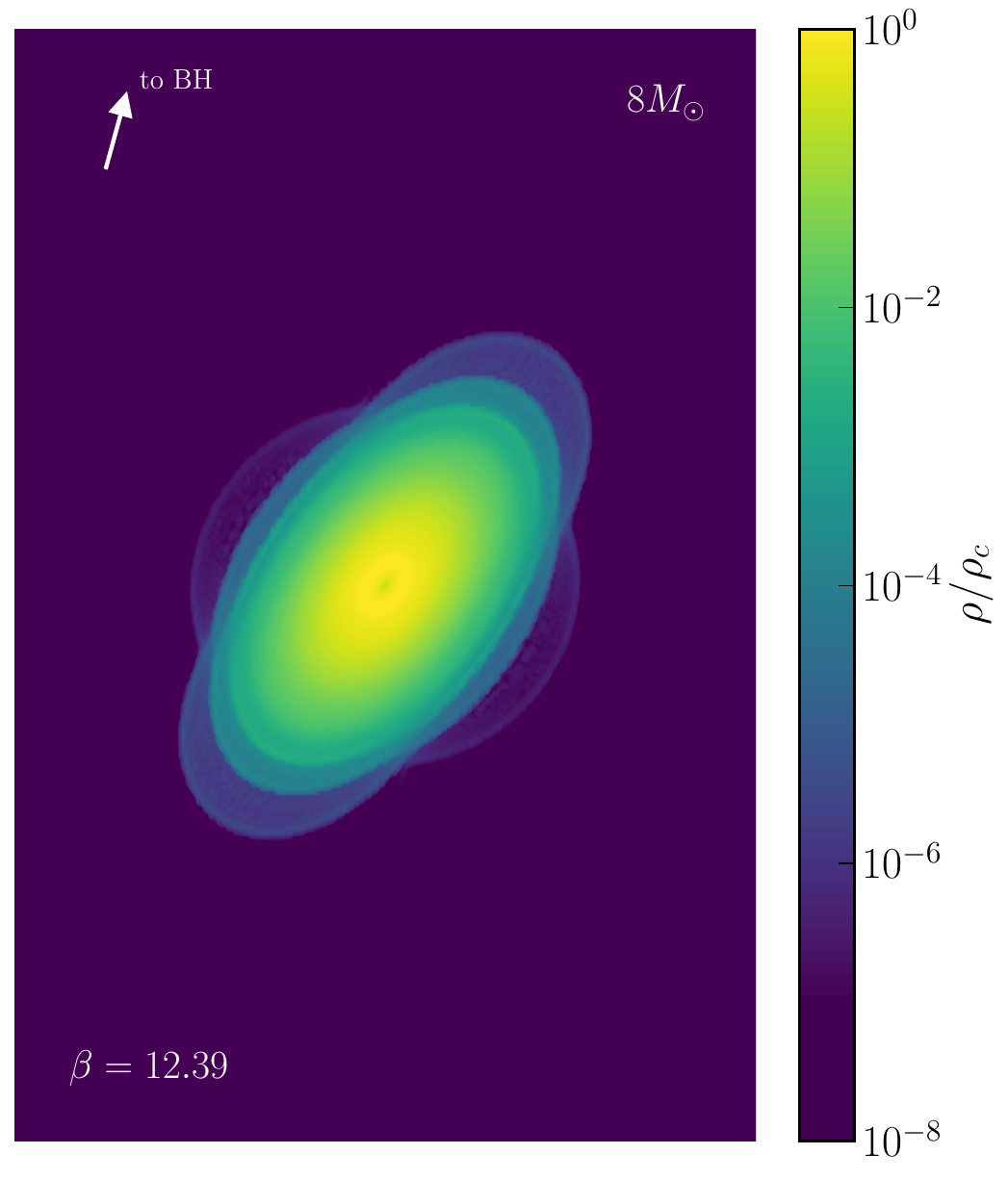}}
\caption{Equatorial density slices at $t=t_{\rm TDE}$ for different values of $M_\star$ and $\beta$. The top row corresponds to $\beta=1.05\beta_{\rm crit}$, the middle row to $\beta=1.76\beta_{\rm crit}$ (equivalent to the four cases described in Fig.~\ref{fig:dMdE-dMdT}), the bottom row to $\beta=5\beta_{\rm crit}$. All cases assume $M_{\rm BH}=10^6M_\odot$. In the top row, the disruption time $t_{\rm TDE}$ is \emph{after} pericenter; in the bottow two rows, it is \emph{before} pericenter. Note that the orientation of this figure is rotated counterclockwise by $\SI{90}{\degree}$ with respect to Fig.~\ref{fig:money-figure}. We denote the direction pointing to the BH with an arrow. At pericenter, the arrow points left. Each panel has a width of $4R_\star$ and a height of $6R_\star$, while the logarithmic color scale is normalized in each panel to the maximum star density.}
\label{fig:density-profiles} 
\end{figure*}

As described in Section~\ref{sec:stage-2}, we sample the deformed star with a fine grid, determining the specific orbital energy $E$ and period $T$ of each fluid element. This allows us to determine the energy distribution $\dd M/\dd E$ and mass fallback rate $\dd M/\dd T$ of the fully disrupted star. We show the results in Fig.~\ref{fig:dMdE-dMdT}, for four different stellar masses, normalizing the energy and fallback time by the dimensional estimates (see \cite{Lodato:2008fr})
\begin{equation}
\Delta E=\frac{GM_{\rm BH}R_\star}{R_p^2}~,\qquad \Delta T=\frac{GM_{\rm BH}}{\Delta E^{3/2}}~.
\label{eqn:deltaEdeltaT}
\end{equation}
Performing about $10^8$ samples of the stellar fluid elements, the required computational time is about 1 minute in total.

The energy distributions are wider compared to those of the unperturbed stars. However, they do not feature a central plateau as seen in hydrodynamical simulations \cite{Lodato:2008fr,Law-Smith:2020zkq,Ryu:2020cqv}. This is likely due the fact that we ignore the fluid self-gravity in the second stage of our model (\cite{Coughlin:2016xka,Coughlin:2019pqk}). The fallback rate curves are consistent with previous results in the literature, peaking at $\mathcal O(10\Delta T)$ and asymptoting $\sim T^{-5/3}$ for large $T$. When comparing with the numerical simulations in (\cite{2023MNRAS.526.2323F}), the peak timescales for the fallback rates modeled using our formalism, while being a significant improvement over the frozen-in approximation, appear to be systematically shorter. This is largely due to the abrupt transition from the perturbative stellar oscillation phase to the free-particle regime in the current model, neglecting fluid relaxation driven by pressure gradients and self-gravity.

In Fig.~\ref{fig:density-profiles} we provide a visualization of the deformed stars at $t=t_{\rm TDE}$, for various masses and values of $\beta>\beta_{\rm crit}$. We create the figure by sampling the fluid elements on an equatorial slice of the star (uniformly in the azimuthal angle $\phi$), and computing their Lagrangian displacements as in Eq.~\eqref{eqn:xi} and~\eqref{eq:fluid_amp_sol}. The positions of all fluid elements (weighed by their mass) are then binned to create a density map, to which we add a constant background equal to $10^{-8}\rho_c$ to improve visual clarity. It is important to realize that the disruption time is different for each case, generally being after pericenter in small-$\beta$ cases, and before pericenter for large $\beta$. In the latter case, the stars appear less deformed, but their energy distribution is wider because $r_p$ is smaller. In all cases, the stars appear to be elongated not towards the BH, but instead along a direction similar to their orbital motion. This ``lag'' effect is due to their dynamical tidal deformation, and qualitatively matches what is observed in hydrodynamical simulations immediately before the star's disruption (\cite{1993ApJ...418..163K,1993ApJ...418..181K}).

Low-mass stars are deformed in a more homogeneous way compared to high-mass ones. In the latter case, the core and the envelope are stretched along different directions. In some cases, the outer shells of the envelope are observed to intersect each other, a feature that likely signals the departure from the perturbative regime. It is likely that our linear model of tidal deformation ceases to be a good description in those cases, and that nonlinear corrections may be required to achieve more accurate predictions. Because in large-$\beta$ events the disruption happens when the star has not had enough time to deform significantly, it is plausible that our perturbative approach works better in those cases. Accurate modeling of large-$\beta$ events also requires to adopt relativistic orbits, which can be done within our model in a straightforward way.

\section{Discussion and Conclusions}
\label{sec:con}

The modeling of TDEs in the literature has followed two different approaches so far: on one hand, rough semi-analytical estimates; on the other hand, expensive hydrodynamical simulations. In this paper, we propose a new model that improves the accuracy of common semi-analytical arguments while still remaining computationally efficient. We achieve this by matching two different stages: stellar linear oscillation theory with free-falling particles. We determine the transition between the two stages introducing a novel TDE criterion, which is calibrated by matching the critical impact parameter to the one observed in hydrodynamical simulations available in the literature. This model allows us to quickly compute the stellar energy distribution and the mass fallback rate.

We find that our deformed stellar density profiles and fallback rates are in good qualitative agreement with those found in the literature. The decomposition of the stellar deformation into eigenmodes also allows us to see which types of oscillations ($g$-modes, $f$-mode, or $p$-modes) are more excited during a TDE. We find that the fractional contribution of different types of oscillations is universal, only depending on the stellar mass, and not on the details of the orbit.

In our analysis, we made a number of approximations. First of all, for simplicity and ease of presentation, we considered Newtonian orbits. The extension to the relativistic case is straightforward, as it only requires to integrate Kerr geodesics and replace the Newtonian tidal field $E_{ij}$ with its relativistic generalization $E_{\mu\nu}=C_{\mu\nu\rho\sigma}u^\rho u^\sigma$, where $C_{\mu\nu\rho\sigma}$ is the Weyl tensor and $u^\mu$ is the star's 4-velocity. This can allow one to study high-$\beta$ encounters with better accuracy---a scenario where the linear tide approximation of our model should perform well due to the small stellar deformation at $t=t_{\rm TDE}$.

Nonlinearities and high-order multipoles are expected to play a significant role in low-$\beta$ encounters, presumably changing the disruption point and the density profile. Our model can be modified to include nonlinearities by adding a three-mode coupling to the mode action \eqref{eq:fluid_lag}, see \cite{Wu:2000ys,Weinberg:2011wf}.

The main approximation adopted in the second stage of our model, where the fluid elements are treated as free particles, is neglecting self-gravity. The deeper the encounter, the worse this approximation likely becomes (opposite to the linear tide hypothesis), because the star is more compact at $t=t_{\rm TDE}$. Self-gravity corrections have been studied in the literature, see e.g.\ (\cite{Coughlin:2016xka,Coughlin:2019pqk}).

It would be interesting to apply our approach to related astrophysical phenomena. A setup very similar to the one studied in this paper are partial tidal disruptions, where only the outer layers of the star are stripped off during the close encounter. Modeling this kind of events would certainly require us to generalize our disruption criterion, for example using the maximum value of $E_Q/|U_{\rm bind}|$ to determine the amount of material that is stripped away. In doing so, one could model TDEs by stellar mass binary BHs (\cite{Lopez:2018nkj}). Another appealing scenario are quasi-periodic eruptions (QPEs), where stellar oscillations can be repeatedly excited through close encounters.

We hope that the simplicity of the approach we presented, together with the large amount of possible extensions and applications, can prove useful and insightful to model these complicated astrophysical events.

\emph{Authors' contribution.} ZZ initiated the project, constructed the two-stage model and wrote the code used in Section 2. GMT wrote the code used in Section 3 and produced all the figures. IMR also initiated the project, setup the fluid model for stellar oscillations in Section 2 and contributed to define the TDE criterion. JL contributed through discussions and also contributed to the TDE criterion.

\section*{Acknowledgements}

We thank Shu Yan Lau, Enrico Ramirez-Ruiz, Hang Yu, Matias Zaldarriaga for insightful discussions. GMT acknowledges support from the Sivian Fund at the Institute for Advanced Study. IMR and JL are partially supported by the US Department of Energy (HEP) Award DE-SC0013528.

\vspace{8pt}

The code used in this work is available on GitHub (\cite{PerTDE}).

\bibliography{bib}

\end{document}